\documentclass[nonacm, screen]{acmart}
\usepackage{amsmath}
\usepackage{xcolor}
\usepackage{subcaption} 

\newtheorem{theorem}{Theorem}

\newcommand{\powerOfD}{$d$-way balanced allocation}

\begin{document}
\title{Balanced allocation: considerations from large scale service environments}

\author{Amer Diwan}
\email{diwan@google.com}
\affiliation{%
  \institution{Google}
  \city{Mountain View}
  \state{CA}
  \country{USA}
}

\author{Prabhakar Raghavan}
\email{pragh@google.com}
\affiliation{%
  \institution{Google}
  \city{Mountain View}
  \state{CA}
  \country{USA}
}

\author{Eli Upfal}
\email{eli@cs.brown.edu}
\affiliation{%
  \institution{Brown University}
  \city{Providence}
  \state{RI}
  \country{USA}
}

\begin{abstract}
    We study \powerOfD{}, which assigns each incoming job to the lightest loaded among $d$ randomly chosen servers. While prior work has extensively studied the performance of the basic scheme, there has been less published work on adapting this technique to many aspects of large-scale systems.  Based on our experience in building and running planet-scale cloud applications, we extend the understanding of \powerOfD{} along the following dimensions:

      \begin{enumerate}
        \item \textbf{Bursts}: Events such as breaking news can produce bursts of requests that may temporarily exceed the servicing capacity of the system.  Thus, we explore what happens during a burst and how long it takes for the system to recover from such bursts.
        \item \textbf{Priorities}: Production systems need to handle jobs with a mix of priorities (e.g., user facing requests may be high priority while other requests may be low priority).  We extend \powerOfD{} to handle multiple priorities.
        \item \textbf{Noise}: Production systems are often typically distributed and thus \powerOfD{} must work with stale or incorrect information.  Thus we explore the impact of noisy information and their interactions with bursts and priorities.
    \end{enumerate}
    
We explore the above using both extensive simulations and analytical arguments.  Specifically we show, (i) using simulations, that \powerOfD{} quickly recovers from bursts and can gracefully handle priorities and noise; and (ii) that analysis of the underlying generative models complements our simulations and provides insight into our simulation results.
\end{abstract}
\maketitle

\section{Introduction}
\label{sec:intro}
Planet-scale cloud applications are globally distributed and run on many tens of thousands of servers.  At this scale, effective load balancing is critical for cost and latency.  It is critical for cost because 
poor load balancing effectively reduces the capacity of the system, which requires service providers to over-provision and thus waste resources. It is critical for latency because poor load balancing creates a utilization imbalance across servers and servers with higher utilization have worse latency and failures (such as "out of memory") than servers with lower utilization.  Our experience with building and running planet-scale cloud applications bears this out.  This paper explores \powerOfD{} algorithms under a variety of contexts motivated by large service environments, to deepen our understanding of when and why they are effective.  We use a combination of simulations and mathematical modeling and analysis to derive these insights.

Scheduling for planet-scale cloud applications must handle complex requirements.  First, it must be able to handle sudden bursts of requests.  For example, soon after a celebrity scandal breaks or a natural disaster occurs, Web search may see a burst in queries from users searching for information about the celebrity or locale.  These bursts may temporarily exceed the serving capacity of the system.  Thus, a good scheduling strategy should be able to recover from the burst, ideally as quickly as possible.
Second, it must handle requests of different priorities; it is common for the load to contain a mix of priorities, perhaps using high priority for user-facing requests, medium priority for background but essential work (e.g., for sending notifications to users on breaking news events), and low priority for deferrable work (such as re-indexing and model retraining).  A good scheduling strategy should minimize any negative impact on the highest priority requests even during bursts; yet it should let the lower priority requests soak up the remaining capacity.  Finally, it must handle imperfect information.  For example, when balancing requests across globally-distributed servers, the load balancer cannot have perfect information about every server; its information may be stale.  These requirements mean that scheduling must be robust: it should behave well in the steady state even with imperfect information and should revert to steady state quickly after a burst. 

We explore the applications of \powerOfD{} in the context of these problems. We use 
both extensive simulations and analytical arguments.  Our simulations demonstrate the advantage of \powerOfD{} in recovering from bursts,  handling priorities, and dealing with noisy and stale information. The accompanying  analysis provides insight into our simulation results. In particular, it proves that the double exponential decay of queue depth distribution --- the primary mathematical characteristic of 
\powerOfD{} --- still holds in the variety of settings (bursts, priorities, and noise) arising in large-scale cloud services. 

The rest of the paper is organized as follows.   Section~\ref{sec:framework}  details our model, methodology and metrics, aligning as far as possible with classical models, yet also adapting to a setting where we inject a load into the system that exceeds capacity (when handling bursts). All our studies --- both theory and simulations --- vary $d$, the number of \textit{probes} made by the scheduler in deciding which queue to assign an incoming job to. The case $d=1$ corresponds to randomly selecting a queue and is used in production; for example both Gmail and Dynamo use consistent hashing to pick the server for the next user or transaction~\cite{ardelean2018performance,decandia2007dynamo}.  Higher $d$ are also common in production systems~\cite{nginx_power_of_two}.

\paragraph{\bf Baseline:} Section ~\ref{sec:simple} presents a simulation of \powerOfD{}\ without bursts, priorities, or noise; it also summarizes classical analysis for this simple setting. The content of this section is not novel; rather, we present it as a baseline for the variants of \powerOfD{}\ to follow.

\paragraph{\bf Recovering from bursts:} Section~\ref{sec:bursts} focuses on subjecting the system to one or more bursts during which the arrival rate exceeds the system capacity. Of interest is the rate at which the system recovers for various values of $d$. Simulations reveal that $d=1$ is slow to recover whereas  $d>1$ recovers rapidly. We complement this with a theoretical analysis that proves a significant asymptotic gap between the recovery times of systems with $d=1$ and $d \geq 2$.

\paragraph{\bf Queues with priorities:} Section~\ref{sec:priorities} explores the setting where each arriving job comes with one of three priorities. We simulate four plausible priority schemes for the scheduler; these schemes differ in how the scheduler picks between the $d$ queues. A queue may contain a mixture of job priorities, each with its own ``depth''. Our results show that jobs of the top priority fare relatively well for a variety of values of $d$ while lower priority jobs fare worse, particularly under $d=1$.
We complement this finding with a theoretical analysis that gives the exact distribution of the number of queued items of each priority, as a function of arrival rate and $d$. Our simulation results mirror the analytically predicted distribution.

\paragraph{\bf Stale and noisy information:} Section~\ref{sec:fuzz} explores the reality that in large scale production services, exact knowledge of all queue depths is unavailable. We consider two models of the impact of poor information on queue depths. In the first model the scheduler uses queue depths that are stale by a temporal \textit{lag} parameter, to capture delays in the propagation of depth information. Building on prior work~\cite{dahlin2000interpreting,mitzenmacher2000useful} we conduct in-depth simulations to capture the phenomenon that under such laggard information, larger values of $d$ may actually lead to poorer results. We provide (for several system loads) simulations suggesting the ``break points'' at which it becomes profitable to drop to smaller values of $d$. In our second model, if the queue depths being compared are within a parameter $b$ (which we call \textit{fuzz}) we declare a random winner between them. Simulations show that larger values of $d$ are more resilient to this model of noise. We complement these with an analysis that characterizes the distribution as a function of $b$ (as well as the other model parameters). We note in this context that: (a) several other error models can be viewed as convex combinations of fuzz and thus their distribution functions can be derived from our analysis; (b) in a trivial error model where every scheduler comparison is faulty with constant probability, the queue depth distribution is within a constant factor of that for Random ($d=1$).

Section~\ref{sec:all} evaluates a combination of bursts, priorities, and imperfect information to explore their interactions.
Section~\ref{sec:related} reviews related work in more detail, in the context of the above contributions.

\section{Framework}
\label{sec:framework}

The remainder of this paper explores Random and \powerOfD{} algorithms using simulations and rigorous analysis.  \powerOfD{} samples $d$ queues when scheduling a new job and picks the queue with the fewest queued jobs. This section describes the setting in which we explore variations of \powerOfD.

\subsection{The core model}
We consider a system of $n$ queues. New jobs arrive according to a Poisson process with rate $\lambda$. Jobs have independent service time distributed 
with parameter $\mu/n$. With $n$ queues, the rate of departure from the system is $\mu$, and we assume $1 > \mu > \lambda>0$, so the system is stable.  All of our analyses are based on this model.

\subsection{Simulator}
To efficiently simulate a large number of queues we use an event driven simulator. 
Our discrete time simulator operates in a series of \textit{jumps}; each jump is equiprobably an \textit{arrival} or a \textit{departure} jump. To correctly compare the various scenarios we designed the simulator so that the inter-jump distribution does not change in time and is independent of the parameters of the model.
An arrival jump generates one or more jobs that are sequentially despatched in time. Because in one of our settings we have arrival \textit{bursts} --- where the arrival rate can temporarily exceed system capacity --- it is necessary for us to model the possibility of more than a single job being generated at a jump. To this end we model the number of arriving jobs in an arrival jump as being Poisson distributed with parameter $\lambda$. A departure jump picks one of the $n$ queues uniformly at random and with probability $\mu$ retires the highest priority job in that queue (if the queue is nonempty). We will apply and extend this simulator to study temporary bursts in arrivals (Section~\ref{sec:bursts}), multiple priorities in Section~\ref{sec:priorities} (where each queue processes jobs of higher priority before jobs with lower priorities), as well as to the setting where information on the load of a queue is noisy (Section~\ref{sec:fuzz}).

To verify that this discrete time process faithfully simulates the core model, suppose that $n+3$ independent continuous-time  processes are running simultaneously. The times to the next event of the processes are exponentially distributed with the following parameters: $\lambda$ for the true arrival process, $1-\lambda$ for a "dummy" arrival process, $\mu/n$ for next termination in each of the $n$ queues (even if they are empty), and $1-\mu$ for the last "dummy" process. The sum of the parameters is 2, so the inter-jump distribution is exponential with parameter 1/2, while the events corresponding to true arrivals or departures follow the original Poisson process.
The probabilities of (true) arrival or departure in a simulator jump are the same as the probabilities of these events in the corresponding jump in the continuous process, which follows the original process. In particular, the ratio between the probability of a true departure and the probability of a true arrival in a jump is
$\frac{r\mu}{n}/(\lambda+\frac{r\mu}{n})$ vs. $\lambda/(\lambda+\frac{r\mu}{n})$, where $r$ is the number of non-empty queues at that time (even when the simulator does not keep track of the number of non-empty queues).

We use simulations rather than production measurements in this paper so that we can efficiently explore a wide space of configurations. 

\subsection{Simulation details}
\label{sec:simdet}
For ease of exploration and exposition, we partition our system into three roles: \textit{load manager}, \textit{scheduler}, and \textit{queues}. The \textit{load manager} provides a sequence of arrival or departure jumps as described above. A \textit{scheduler} picks a queue for each arriving job using \powerOfD. Unless we indicate otherwise, $d$ ranges over the values $1,2,3$ and $4$; recall that $d=1$ corresponds to Random allocation.
Our simulations use $n=1000$ queues (i.e., $q=1000$) and $\lambda= 0.95$ except when we indicate otherwise\footnote{The Appendix has data for $\lambda=0.75$ which represents another interesting point in the space where both the steady state and bursty load are below system capacity; since the conclusions are similar, we omit them from the main body of the paper.}.  Planet-scale production systems often use tens to hundreds of thousands of servers but for scalability they break down the load balancing problem into two levels (e.g., the first level may pick the cell or metro or continent and the second level picks the actual server).  Thus $q=1000$ is appropriate particularly for the second level of load balancing. 

Each simulation runs for 12 million jumps.  $\lambda = 0.95$ ensures that our queues are heavily utilized (i.e., 95\%) and thus our simulation exhibits interesting behavior.   

We inspect the state of all 1000 queues at the end of each simulation or just before a burst (to measure the state of recovery from any prior bursts, see Section~\ref{sec:bursts}).  At each inspection we record the average and maximum number of jobs pending in each queue (the queue depth) and use these in our findings.  

We use both plots and tables to illustrate our simulation results. To expose the variance within a simulation, we show a \textit{single} simulation run in each plot. However to maintain statistical significance, each table summarizes the average of measurements from 30 simulation runs; thus the tables are the definitive source of our conclusions.



\section{Baseline: $d$-way balanced allocation in a steady state with no priorities and no errors.}
\label{sec:simple}
As a baseline to compare with our results in Sections \ref{sec:bursts} - \ref{sec:fuzz}, this section presents a simulation as well as known analytical results about the "supermarket" model:
\powerOfD~ in a steady state with no bursts, job priorities or probing errors. 

\subsection{Simulation results}

\begin{figure}[htb]
    \centering
    \begin{minipage}[b]{0.55\textwidth}
        \centering
        \includegraphics[width=\textwidth]{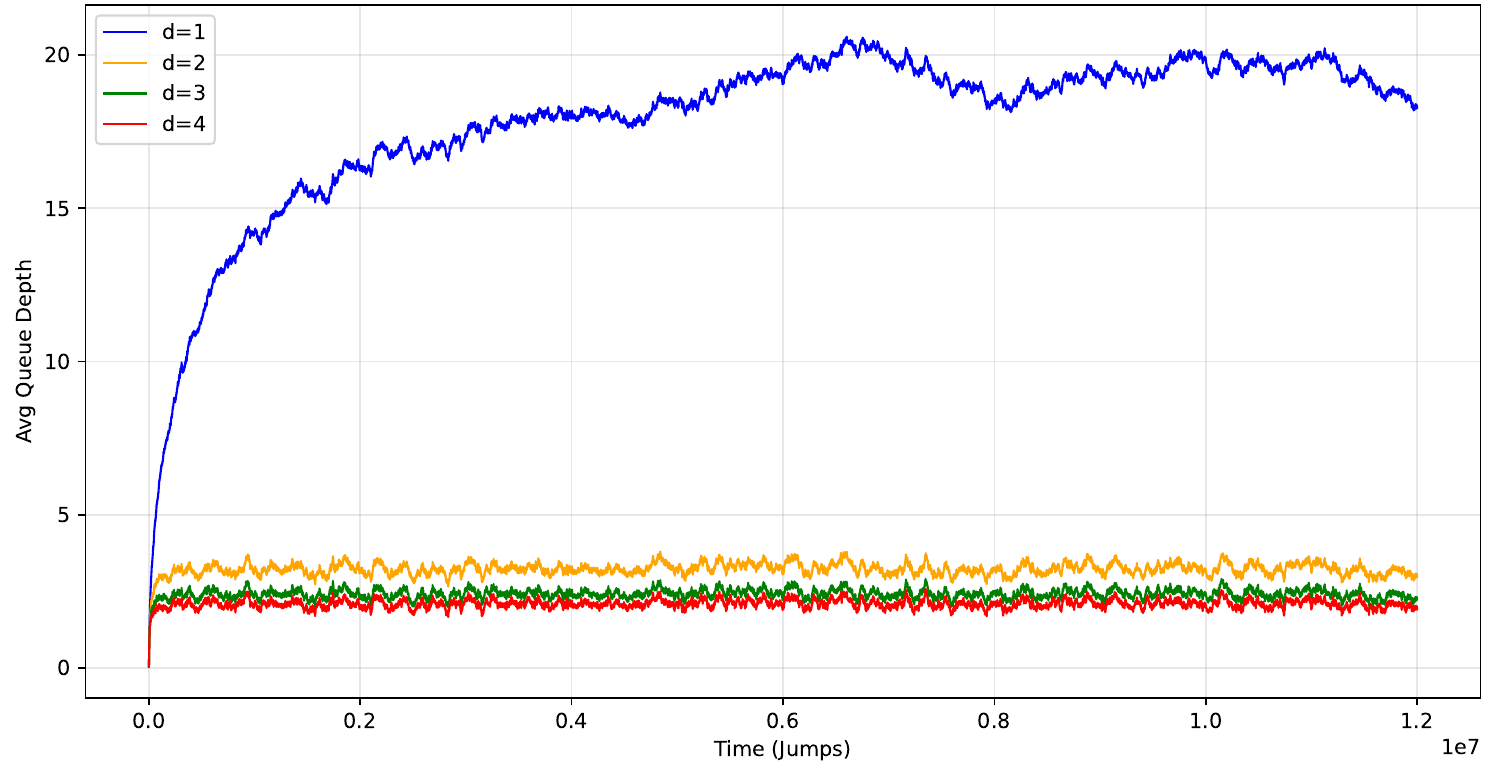}
        \caption{Time series for the baseline configuration.} 
        \label{fig:simple-ts}
    \end{minipage}\hfill
    \begin{minipage}[b]{0.40\textwidth}
        \centering
        \begin{tabular}{l|rr}
            \toprule
            $d$ & Avg Depth & Max Depth \\
            \midrule
            1 & 18.75 & 136.30 \\
            2 & 3.24 & 6.77 \\
            3 & 2.40 & 4.70 \\
            4 & 2.11 & 4.00 \\
            \bottomrule
        \end{tabular}
        \vspace{10pt} 
        \captionof{table}{Queue Depth Statistics by $d$}
        \label{tab:simplest}
    \end{minipage}
\end{figure}

Figure~\ref{fig:simple-ts} shows the average number of jobs in a queue over the course of a single simulation. 
We see that each configuration with $d>1$ stabilizes soon after the start of the simulation (we can not even make out a ramp up for $d>1$), whereas  $d=1$ stabilizes much further into the simulation.

Table~\ref{tab:simplest} summarizes Figure~\ref{fig:simple-ts} across 30 simulations by averaging queue depths in our measurements, as outlined in Section~\ref{sec:simdet}.
The ``Avg Depth'' column shows that the average number of jobs in a queue drops sharply from 18.75 ($d=1$) to 3.24 ($d=2$).  This means that each job must wait for (on average) 17.75 jobs ahead of it to depart with $d=1$ but only for 2.24 jobs with $d=2$. Further increasing $d$ to 4 offers an additional improvement of 50\% (dropping from waiting for 2.24 jobs to waiting for 1.11 jobs).  The ``Max Depth'' column gives the maximum number of jobs in any queue in a measurement, averaged over 30 simulations.  We see that the maximum depth for $d=1$ is 136.3 compared to 6.77 for $d=2$.  The relatively tight distribution for $d>1$ illustrates how well \powerOfD{} balances load across queues.  Higher values of $d$ perform better than $d=2$ and thus if it is affordable, using more than $d=2$ is worthwhile.

\subsection{Analytical results (based on prior work)}
The steady state properties of the \powerOfD~ (for $d>1$) with a
Poisson arrival rate $\lambda<1$ and exponential service time with parameter $\mu>\lambda$, has been studied in ~\cite{vvedenskaya1996queueing,963420}. It was shown that $\pi_i$, the probability that a queue has $\geq i$ jobs, converges to an invariant distribution which has a double exponential decay in $i$: 
$$\pi_0 = (1-\lambda) ~~~\mbox{and} ~~~~\pi_i =\lambda^{\frac{d^i-1}{d-1}}.$$
There is no simple closed form for the expected steady state queue size. However, the probabilities decrease rapidly so the first 4-5 terms in 
 $\sum_{i\geq 1} \pi_i =\sum_{i\geq 1} \lambda^{\frac{d^i-1}{d-1}}$ suffice for a good approximation.

\section{Incorporating bursts}
\label{sec:bursts}

Motivated by the spiky events we have described in Section~\ref{sec:intro}, we now explore how bursts affect \powerOfD. A burst  temporarily magnifies the arrival rate beyond system capacity, leading to increased queue lengths. We study whether and how quickly a scheduler clears the backlog after a burst.
We model a burst by modifying the Poisson variable in an arrival jump to have expectation $\lambda_b =c\lambda$ that can be greater than 1, before returning to the ``ambient'' $\lambda<1$. 
Our simulations and mathematical analysis demonstrate and quantify the sharp gap between Random and Balanced allocation schedules in the speed of recovery from a burst.

\subsection{Simulation results}
\label{sec:burstsim}
Each burst lasts for 2\% of the total simulation during which we increase $\lambda$ by a factor of $1.2$x.  Given this, we can calculate the minimum recovery time from a burst that lasts 1\% of the simulation:
\begin{equation}
(\lambda * 1.2 - 1) / (1 - \lambda) .
\end{equation}

Plugging in $\lambda =0.95$, we get 2.8 for this value; thus if a burst lasts 2\% of the simulation, it will take us $2.8 * 2 = 5.6$\% of the duration of the simulation to recover with a scheduler that has perfect knowledge of all queue depths at all times; thus it drains the accumulated jobs with perfect efficiency within an \textit{optimal recovery period}.
Thus a simulation run consists of periodic cycles each consisting of (1) a burst; (2) an optimal recovery period and finally (3) a buffer period at the end of which we tabulate metrics; then the cycle begins again with a new burst. For comparison we also report data for a simulation with zero bursts.

We inject between zero and four equidistant bursts over the in a simulation, after skipping the first 60\% of the simulation to enable the system to warm up and stabilize.  Such back-to-back bursts are common in production: e.g., from users checking election results as they get announced.

\begin{figure}[htb]
    \centering
    \includegraphics[width=0.6\textwidth]{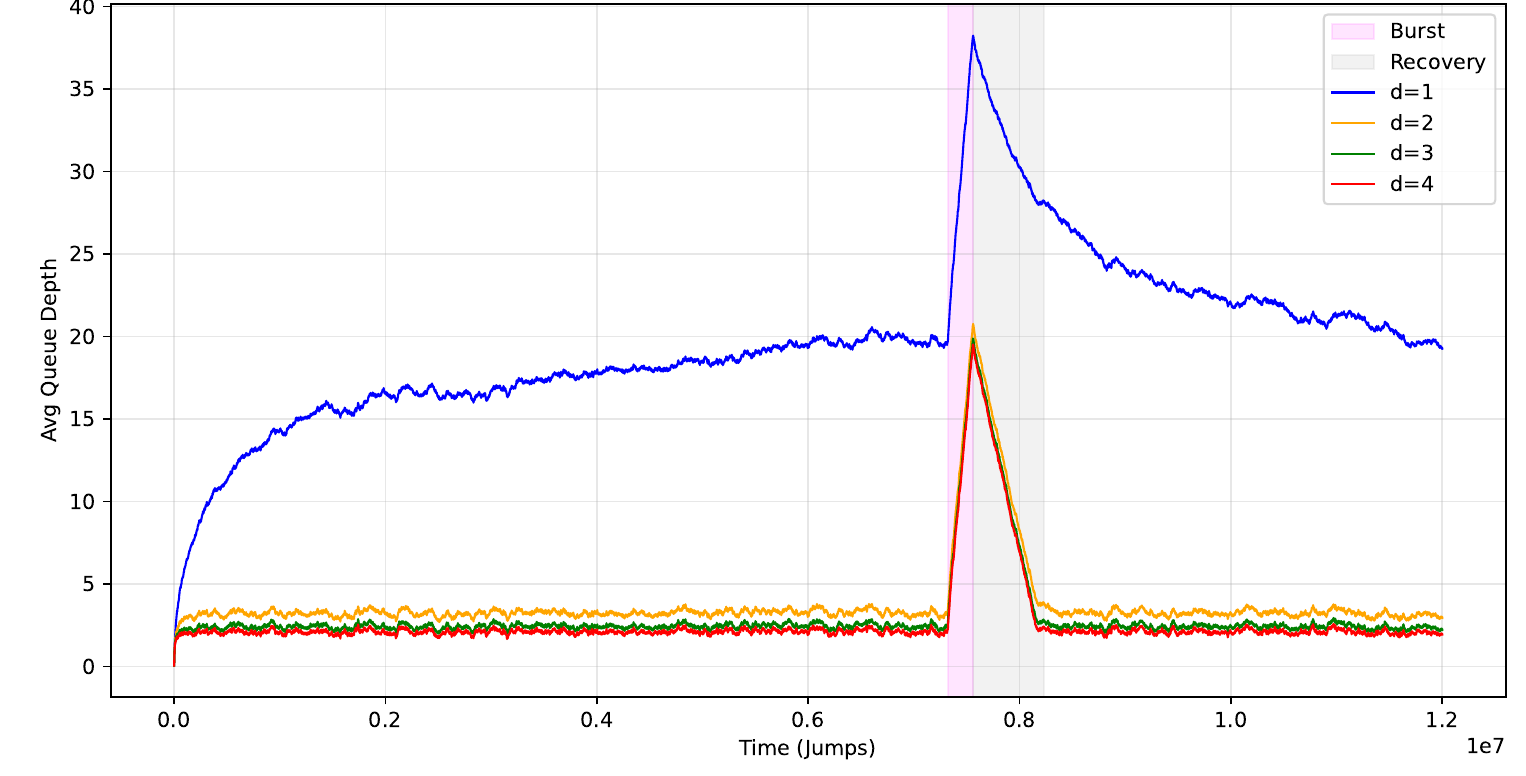}
    \caption{Queue sizes with 1 burst.} 
    \label{fig:burst1}
\end{figure}

Figure~\ref{fig:burst1} shows what happens when we insert a single burst after the warmup (60\% into the simulation).  The magenta area marks the burst, and the gray area after the magenta marks the optimal recovery period.  We see that $d>1$ recover to the state before the burst soon after the burst, but $d=1$ takes the rest of the simulation to get to the state before the burst.

\begin{figure}[htb]
    \centering
    \includegraphics[width=0.6\textwidth]{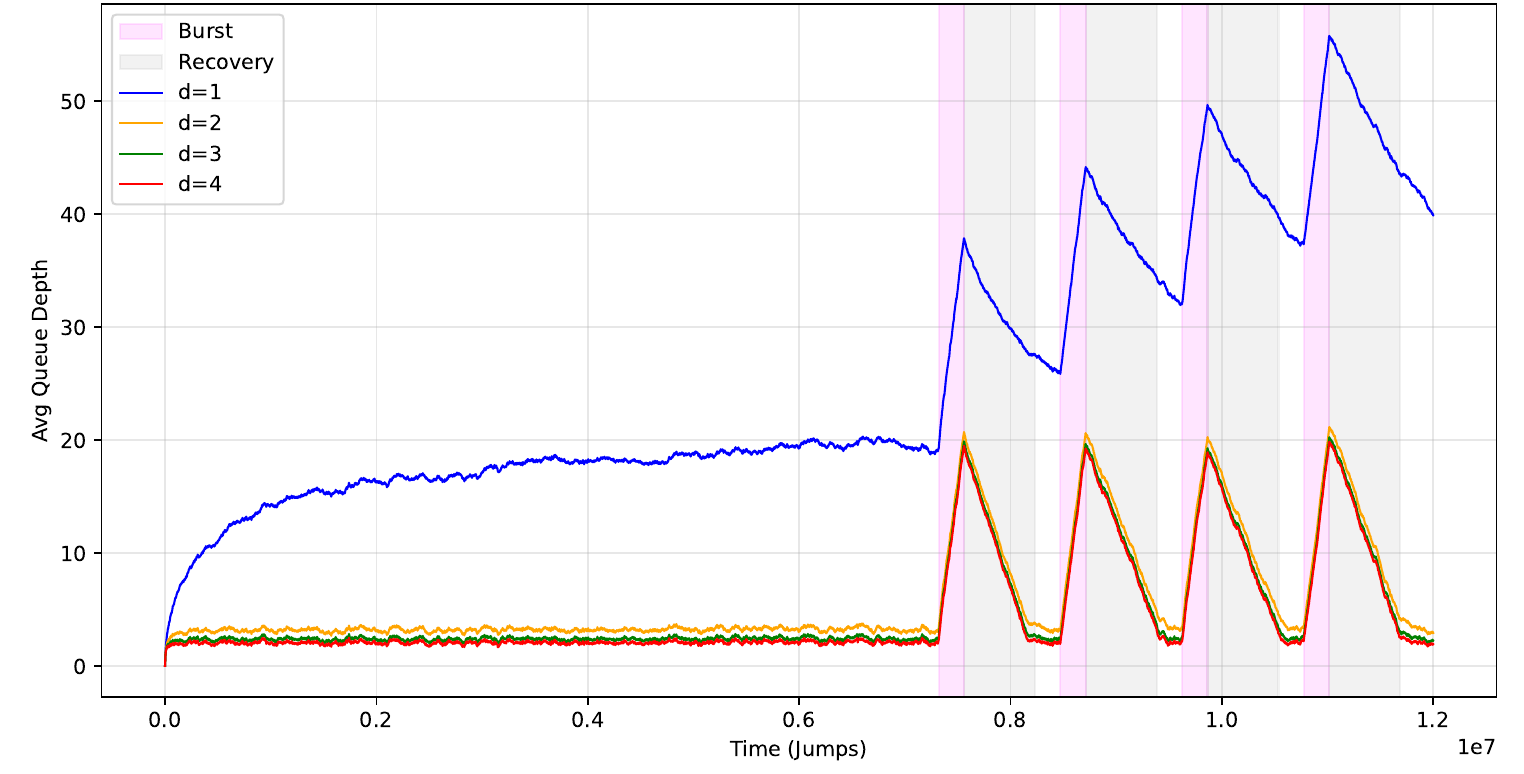}
    \caption{Queue depth with 4 bursts.} 
    \label{fig:burst}
\end{figure}

Figure~\ref{fig:burst} illustrates a more complex situation with 4 bursts and each curve gives the average queue depth for a particular value of $d$.  The white areas between the bursts mark the buffer periods, at the end of each of which we measure and summarize all 1000 queue depths.
As we add more bursts, the system has less and less time to recover from a burst before the next burst arrives.  If the system is not able to recover, we will see jobs accumulate at queues with each subsequent burst.
 
\begin{table}
\begin{subtable}[t]{0.45\textwidth}
\caption{Average}
\begin{tabular}{l|rrrr}
\toprule
 & \multicolumn{4}{c}{Burst Count} \\
$d$ & 0 & 2 & 3 & 4 \\
\midrule
1 & 18.91 & 26.86 & 30.23 & 32.23 \\
2 & 3.26 & 3.20 & 3.25 & 3.24 \\
3 & 2.41 & 2.46 & 2.42 & 2.42 \\
4 & 2.06 & 2.14 & 2.08 & 2.10 \\
\bottomrule
\end{tabular}
\end{subtable}
\hfill
\begin{subtable}[t]{0.45\textwidth}
\caption{Maximum}
\begin{tabular}{l|rrrr}
\toprule
 & \multicolumn{4}{c}{Burst Count} \\
$d$ & 0 & 2 & 3 & 4 \\
\midrule
1 & 143.53 & 169.83 & 194.33 & 209.87 \\
2 & 6.77 & 6.70 & 6.97 & 7.03 \\
3 & 4.80 & 4.80 & 4.97 & 4.97 \\
4 & 4.00 & 4.00 & 4.00 & 4.00 \\
\bottomrule
\end{tabular}
\end{subtable}
\caption{Queue depth for a combination of $d$ (rows) and number of bursts (columns)}
\label{tab:burst-avg-and-max}
\end{table}

Tables~\ref{tab:burst-avg-and-max} (a) and (b) summarize how well \powerOfD{} handles bursts with increasing $d$.  It gives the average and maximum queue depth across the measurements and averaged over all 30 simulations.  For Random (i.e., $d=1$) we see that average and maximum queue depth becomes progressively worse as we increase the number of bursts and thus decrease the spacing between bursts.  The average queue depth increases by 70\% as we go from 0 bursts to 4; the maximum queue depth increases by 46\%.  Thus, adding bursts to Random degrades the distribution of queue lengths which in turn translates into higher latency.

Figure~\ref{fig:burst} suggests that \powerOfD{} recovers between bursts when we have 4 bursts (in the single simulation plotted here). From the tables we see that $d=4$ recovers between bursts even with 4 bursts: maximum and average queue depths hardly change as we reduce the distance between bursts by increasing the burst count.  For $d=2$ and $d=3$ we see a small increase in maximum queue depth as we increase the burst count indicating that it may still be recovering at the end of the buffer period.

To quantify how long it takes to recover we used a sliding window of 10000 jumps in the period(s) following optimal recovery period(s), to see when the average queue depth drops to the level just before the first burst.  We found that $d=4$ recovers within the first such window.

Qualitatively, \powerOfD, particularly with $d=4$ recovers rapidly from bursts and is nearly optimal in doing so. Random, in contrast, does not recover even when the bursts are far apart; this is because Random continues to send requests to loaded queues while $d>1$ attempts to avoid those queues.

\subsection{Analytical results}

The key to understanding the huge gap in recovery time between Random and $d$-way balanced allocation is a  
phenomenon first analyzed for the static case (only arrivals, no servicing/retiring of jobs) in the celebrated "heavily loaded" paper ~\cite{10.1145/335305.335411}. The paper analyzes a combinatorial process that sequentially places $m>>n$ balls (corresponding to jobs) in $n$ bins (corresponding to servers) using 2-way balanced allocation. A major result of the paper is that with high probability the maximum bin load after placing the $m$ balls is $m/n+O(\log\log n)$. Thus, surprisingly,
the deviation from the expectation is a function of $n$ but not $m$. In contrast, when $m$ balls are placed by Random in $n$ bins, the standard deviation of the number of balls in a bin is $\Theta(\sqrt{m/n})$. This implies that the expected number of queues with at least $m/n+c\sqrt{m/n}$ is $\Omega(n)$.\footnote{This relies on a lower bound on the upper tail of the binomial that can be shown, for instance, by invoking the Berry-Essen theorem~\cite{Feller1971}[XVI.5]}

Suppose that after the $m$ balls are sequentially placed in the $n$ bins we start a removal process
that in a round-robin sequence inspects one bin in each step and removes one ball from the bin if it is non-empty. In 2-way balanced allocation, all
bins will be empty after $m+O(n\log\log n)$ steps,  whereas under Random, we will still have $\Omega(n)$
balls in bins after $m+\Omega(n\sqrt{m/n})$ steps.  

Adapting this observation to the dynamic version with random arrivals and departures is not straightforward. The main difficulty is that while new jobs are placed according to $d$-way balanced allocation, departures occur in queues chosen uniformly at random. Consequently, when the system is not in equilibrium, the distribution of queue depths may not follow a double exponential decay. 

Consider two systems, each with $n$ queues, operating with identical arrival and departure parameters. One system is running $2$-way balanced allocation and the other system is running Random.   Let $\lambda_\text{high} > \mu $, and $0<\lambda_\text{low} <\mu < 1$. The system is empty at jump $0$. In the "burst" interval $(0,T]$, for $T=O(n^4)$, the arrival rate is  $\lambda_{high}$. After jump $T$ the arrival rate is $\lambda_\text{low}$. Denote by $m(T)=T(\lambda_{high}-\mu)+O(1)$ the average queue load at time $T$ in the systems at the end of the burst. (The average queue length in the two systems differs by $O(1)$). We say that a queue recovers from the burst when the distribution of its depth converges to its steady state distribution with parameters $\lambda_\text{low}$ and $\mu$.
Regardless of the scheduler (whether $2$-way allocation or Random), the optimal recovery period (as noted in Section~\ref{sec:burstsim} above) is lower bounded by $n m(T)/(\mu-\lambda_\text{low})$.
We prove bounds on the \textit{incremental} number of jumps for a queue to recover under each scheduler, \textit{beyond the optimal recovery period}, to show that Random is asymptotically slower to recover (with a lower bound) than \powerOfD{} for any $d>2$ (with upper bounds).
\begin{theorem}


\begin{enumerate}
    \item 
    The expected incremental number of jumps to recovery under Random is $\Omega(n{\sqrt{m(T)})}/{(\mu-\lambda_\text{low}}\ )$.
     \item The expected incremental number of jumps to recovery under $2$-way balanced allocation of a given queue is $O(n{\log\log n)}/({\mu-\lambda_\text{low}})$.
     \item 
     The expected incremental number of jumps until the $2$-way balanced allocation system converges to its steady state is $O(n{\log n)}/({\mu-\lambda_\text{low}})$.
\end{enumerate}
\end{theorem}
A similar theorem holds when the systems start in their steady state distribution with parameters $\lambda_\text{low}$ and $\mu$, but in that case the Random system has $\Omega(n)$ more jobs prior to the burst.  We also note that the requirement $T=O(n^4)$ is an artifact of the analysis in ~\cite{berenbrink2025ballsbinsinfiniteprocess}, higher exponents could probably work in our application. The theorem is stated in terms of the recovery of an individual queue. We obtain similar results for system (i.e., all $n$ queues) recovery by adding an $O(\log n)$ factor to the two bounds.

\begin{proof}
(1) We analyze the system running Random as a set of $n$ independent $M/M/1$ queues, each with arrival rate $\lambda_\text{low}/n$ and service rate $\mu/n$. Note that for an even comparison, a step of a queue should count as $n$ jumps of the system.

The average queue depth at time $T$ is  $$T\left(\frac{\lambda_{high}}{n}-\frac{\mu}{n}\right) = m(T)-O(1),$$ and the variance is $$T\left(\frac{\lambda_{high}}{n}(1-\frac{\lambda_{high}}{n})+\frac{\mu}{n}(1-\frac{\mu}{n})\right)\geq m(T).$$
Using the Berry-Essen theorem~\cite{Feller1971}[XVI.5],
$\Omega(n)$ of the queues have depth $\Omega(m(t)+\sqrt{m(t)})$ at time $T$, and each of these queues converges to steady state in expected time $$\Omega\left (\frac{m(t)+\sqrt{m(t)}}{\frac{\mu}{n}-\frac{\lambda_\text{low}}{n}} \right  )=
\Omega\left (\frac{n(m(t)+\sqrt{m(t)})}{\mu-{\lambda_\text{low}}}  \right ),$$ leading to an incremental time to converge of $\Omega(n{\sqrt{m(T)})}/{(\mu-\lambda_\text{low}}\ )$.

(2) We adapt a theorem from ~\cite{berenbrink2025ballsbinsinfiniteprocess} to characterize the state of $2$-way balanced allocation at time $T$.
The discrete time model in ~\cite{berenbrink2025ballsbinsinfiniteprocess} is defined by a sequence $\{\beta(t)\}_{t\geq 1}$, such that at time $t$ with probability $\beta(t)$ the system places a new arrival and otherwise (with probability $(1-\beta(t)$) it deletes an item from a random non-empty queue. Thus, a step in their model corresponds to an expected $2/(\lambda+\mu)$ jumps in our simulation, and an interval with that expected length in our core model. More problematic is that in a departure step their model chooses a queue at random from the set of non-empty queues. In the queuing context, this  implies that the execution time of a job is a function of the number of empty queues. Correcting for these differences, we verify that the execution during the interval $(0,T]$ satisfies their definition of c-good interval (Definition 3.1), and we can apply their Theorem 3.1 to prove that at time $T$, no queue has more than $m(T)+\log\log n$ jobs.

   
   Next, we develop a simple coupling technique to evaluate the expected convergence time of a queue conditioned on its depth at time $T$.

   Let $X_1,\dots,X_n$ denote the $n$ queues of our system and let $Y_1,\dots,Y_n$ denote the queues of a second system which also runs a 2-way balanced allocation algorithm and is in a steady state distribution for parameters $\lambda_\text{low}$, and $\mu$.
   
   Let $x(t)=(x_1(t),\dots,x_n(t))$ be the number of jobs in queues $X_1,\dots,X_n$ at jump $t\geq T$, and
   $y(t)=(y_1(t),\dots,y_n(t))$ be the number of jobs in queues $Y_1,\dots,Y_n$ at jump $t$. Note that $\sum_{i=1}^n x_i(T) =O(nm(T))$, while $\sum_{i=1}^n y_i(T) = O(n)$. 

   To initiate the coupling at time $T$, for $1\leq i\leq n$, color $y_i(T)$ jobs in queue $X_i$ red, the rest are colored white. If $x_i(T)<y_i(T)$, add $y_i(T)-x_i(T)=O(\log\log n)$ new red jobs to queue $X_i$. The extra jobs will have negligible effect on the  convergence time.
   
   The arrival and departure jumps use the same indices in both systems. Suppose that an arrival at time $t$ probes queues $i$ and $j$. If $x_i(t)\geq x_j(t)$ and $y_i(t)\geq y_j(t)$ a new job is added to $Y_j$ and a new red job is added to $X_j$. If  $x_i(t)\geq x_j(t)$ but $y_i(t)< y_j(t)$, a new job is added to $Y_i$, a white job in $X_i$ is colored red (there must be a white job in $X_i$ because $x_i (t)\geq x_j(t)\geq y_j(t) >Y_i(t)$), and a new white job is added to $X_j$. If a departure step at time $t$ chooses $j$ and $y_j(t)>0$ then a job is removed from $Y_j$ and a red job is removed from $X_j$. If $y_j(t)=0$ and $x_j(t)>0$ then a white job is removed from $X_j$.
   
   It is easy to verify that at any time $t\geq t_0$ and for any $1\leq i\leq n$, $X_i$ has exactly $y_i(t)$ red jobs. Also, the total number of white jobs, and the maximum number of white jobs in all queues are monotone non-increasing in $t$. It remains to estimate the rate in which white jobs leave the system. 
   
   The $y(t)$ vector is distributed according to the steady state, therefore the expected number of empty queues at any given jump is $1-\lambda_\text{low}/{\mu}$. Therefore, the expected number of jumps between successive deletions of white jobs in a queue with white jobs  is $n\mu(1-\lambda_\text{low}/{\mu})$. Since no queue has more than $m(T)+O(\log\log n)$ jobs at time $T$, the expected number of jumps until a given queue has no white jobs is 
   $n(m(T)+O(n\log\log n))/(\mu-\lambda_\text{low})$.

   
  (3) When at least one queue has no white jobs, by Theorem 2.1 in~\cite{berenbrink2025ballsbinsinfiniteprocess} the maximum number of jobs in any queue in the system is $O(\log n)$. Thus, in another expected $O(n\log n)$ jumps there are no white jobs in the system and the coupling coalesces.
\end{proof}

\paragraph{Significance:} The analytical results explain the stark contrast in recovery times observed between $d=1$ and $d>1$ in the simulations (Section 4.1). Theorem 1 formally quantifies this "recovery gap," proving that while random allocation ($d=1$) requires a significant incremental recovery period 
($\Omega(n\sqrt{m(t)})$), 2-way balanced allocation converges to its steady state in just $O(n\log  n)$ incremental jumps. This explains why, in Figure 2, the $d>1$ configurations return to their baseline levels almost immediately after the optimal recovery period, whereas the $d=1$ system remains heavily backlogged for the remainder of the simulation. Essentially, the simulations demonstrate that d-way allocation avoids the high-variance "laggard" queues that Theorem 1 identifies as the primary cause of slow recovery in random systems.

\section{Jobs with Priorities}
\label{sec:priorities}

This section explores how to incorporate multiple priorities into \powerOfD.  Most production systems must handle jobs of varying priorities: there is always work that must be done as soon as possible and other work that is less critical.  In this paper we assume a simple model: we always pick the highest priority job in a queue to execute next.  Prior work has explored alternative models (e.g., Stride Scheduling ~\cite{waldspurger1995stride}) that guarantee fairness: i.e., high priority work cannot completely starve out lower priority work.  Niu \textit{et al.}~\cite{9812880} provide an analysis that assumes that lower priority jobs can be migrated from their assigned server to another when needed; our analysis obviates the need to presume such (expensive in practice) operations.

We explore four options for incorporating multiple priorities into \powerOfD. Each option picks a queue from $d$ choices for a request of priority $p$ as follows:
\begin{enumerate}
    \item \textbf{Independent}: Picks the queue with the fewest jobs of priority $p$, breaking ties at random.
    \item \textbf{MineThenTotal}: Picks the queue with the fewest jobs of priority $p$, breaking ties using the total number of jobs in the $d$ queues.
    \item \textbf{TotalThenMine}: Picks the queue with the lowest total number of items, breaking ties using the number of items of priority $p$.
    \item \textbf{CumulativeThenTotal}: Picks the queue with the fewest jobs of priority $p$ or higher, breaking ties using the total number of queued jobs in the probed queues.
\end{enumerate}

Our simulations show that there is no noticeable difference between the performance of the four strategies, despite the fact that 
the \textit{Independent} strategy uses the least amount of information and is easier to implement. Furthermore, both the simulations and the analysis show that adding priorities does not degrade the total utilization of the system. The difference between the average queue depth in scheduling with and without priorities is negligible, for equivalent total loads. The simulations also demonstrate a significant gap between $d=1$ and $d>1$ for all priorities in all strategies. The mathematical analysis proves that the \textit{Independent} strategy extends the phenomenon of double exponential decay of queue length distribution  to priority queues, and that the strategy takes close to optimal use of the service capacity. 


\subsection{Simulation results}
For our simulations each arrival jump picks equiprobably between three priorities $P0, P1$ and $P2$.  We have explored other mixtures (e.g., ones where 80\% of the requests are highest priority) but did not observe any significant differences in results.  The common theme across all of the mixtures is that $P0$ generally fares better because our system treats $P0$ as the primary work and the other priorities consume the left-over capacity.  
Tables~\ref{tab:priority-avg} and \ref{tab:priority-max} give the average and maximum queue depth for different $d$ and with four strategies; for presentation we use the first word of each strategy's name as the column names in our table.

The highest priority ($P0$) actually does much better than Table~\ref{tab:simplest} even for Random scheduling ($(d=1)$) because from $P0$'s perspective the system is underutilized: $P0$ makes up only $1/3$ of the jobs and each queue always executes them first.
All priorities do well as $d$ increases, and \textit{MineThenTotal} is the best or close to the best overall strategy (but \textit{CumulativeThenTotal} performs similarly).  As we have seen before, increasing $d$ is beneficial across the board but its greatest impact is for $P2$: $d=4$ reduces the average and maximum queue depth to about half of $d=2$ for \textit{MineThenTotal}.
To evaluate the utilization of the system with priority scheduling, note that for all strategies and all values of $d$, the sum of queue depth of the three priority is within $3$ to $4\%$ of the corresponding value for the same $d$ in~Table~\ref{tab:simplest}.

\begin{table}
\begin{tabular}{l|rrr|rrr|rrr|rrr}
\toprule
& \multicolumn{3}{c}{CumulativeThenTotal} & \multicolumn{3}{c}{Independent} & \multicolumn{3}{c}{MineThenTotal} & \multicolumn{3}{c}{TotalThenMine} \\
\cmidrule(lr){2-4} \cmidrule(lr){5-7} \cmidrule(lr){8-10} \cmidrule(lr){11-13}
d & P0 & P1 & P2 & P0 & P1 & P2 & P0 & P1 & P2 & P0 & P1 & P2 \\
\midrule
1 & \textbf{0.46} & \textbf{1.26} & \textbf{17.55} & \textbf{0.46} & \textbf{1.26} & \textbf{17.22} & \textbf{0.47} & \textbf{1.26} & \textbf{17.09} & \textbf{0.45} & \textbf{1.24} & \textbf{17.12} \\
2 & \textbf{0.35} & \textbf{0.63} & 2.54 & 0.40 & 0.71 & \textbf{2.11} & \textbf{0.34} & \textbf{0.63} & 2.42 & 0.39 & 0.68 & \textbf{2.15} \\
3 & \textbf{0.32} & \textbf{0.51} & 1.69 & 0.38 & 0.60 & \textbf{1.45} & \textbf{0.33} & \textbf{0.53} & 1.64 & 0.36 & 0.55 & \textbf{1.51} \\
4 & \textbf{0.32} & \textbf{0.47} & 1.41 & 0.38 & 0.56 & \textbf{1.20} & \textbf{0.32} & \textbf{0.49} & 1.37 & 0.34 & \textbf{0.49} & \textbf{1.24} \\
\bottomrule
\end{tabular}
\caption{Average Queue Depth by $d$ and Priority.  \textbf{} highlights results that are within 5\% of the best}
\label{tab:priority-avg}
\end{table}

\begin{table}
\begin{tabular}{l|rrr|rrr|rrr|rrr}
\toprule
& \multicolumn{3}{c}{CumulativeThenTotal} & \multicolumn{3}{c}{Independent} & \multicolumn{3}{c}{MineThenTotal} & \multicolumn{3}{c}{TotalThenMine} \\
\cmidrule(lr){2-4} \cmidrule(lr){5-7} \cmidrule(lr){8-10} \cmidrule(lr){11-13}
d & P0 & P1 & P2 & P0 & P1 & P2 & P0 & P1 & P2 & P0 & P1 & P2 \\
\midrule
1 & 6.07 & \textbf{14.73} & 148.03 & \textbf{5.77} & \textbf{14.27} & \textbf{138.57} & \textbf{5.77} & 15.03 & \textbf{134.70} & \textbf{5.77} & \textbf{14.17} & \textbf{134.57} \\
2 & \textbf{2.27} & \textbf{3.23} & 6.53 & 3.90 & 4.80 & 5.80 & \textbf{2.37} & \textbf{3.10} & 5.77 & 3.57 & 4.10 & \textbf{5.30} \\
3 & \textbf{2.00} & 2.43 & 4.30 & 3.23 & 3.60 & 4.07 & \textbf{2.00} & \textbf{2.03} & \textbf{3.77} & 2.77 & 3.00 & \textbf{3.80} \\
4 & \textbf{1.93} & \textbf{2.03} & 3.70 & 3.07 & 3.03 & 3.53 & \textbf{1.93} & \textbf{2.00} & \textbf{3.00} & 2.30 & 2.37 & \textbf{3.00} \\
\bottomrule
\end{tabular}
\caption{Maximum Queue Depth by $d$ and Priority.  \textbf{Bold} highlights results that are within 5\% of the best. }
\label{tab:priority-max}
\end{table}

\subsection{Analytical results}
We first focus on the case of two priorities, $P0$ and $P1$. We then extend the result to three priorities.

Assume independent Poisson arrival rates, $\lambda_0$ and $\lambda_1$, and independent exponential service times 1. (Given exponential service times we can assume w.l.o.g. that the expected service time is 1 --- else we can normalize the arrival rates). 

We analyze $d$-way balanced allocation with the Independent scheme for handling priorities:
\begin{itemize}
\item
To place a $P_0$ job  we choose $d_0$ random queues and place the job in the queue with the smallest number of $P_0$ jobs (ties broken arbitrarily).
\item
To place an $P_1$-job we choose $d_1$ random queues and place the job in the queue with the smallest number of $P_1$ jobs (ties broken arbitrarily).
\end{itemize}
Theorem~\ref{thm:priorities} shows that the jobs in each priority class behave as if under (different) $d$-way balanced allocations; in particular jobs of the highest priority ``eat up'' the capacity available to them, but even jobs of the lower priority enjoy a doubly exponential distribution of queue depths albeit with a lower base:
\begin{theorem}\label{thm:priorities}
In the above 2 priorities process, 
\begin{enumerate}
\item 
The number of queues with $i$ or more $P_0$-jobs converges to $$\lambda_0^{\frac{d_0^i-1}{d_0-1}}.$$
\item
The distribution of the number of $P_1$-jobs in queues follows the balanced allocation process with arrival rate $\frac{\lambda_1}{1- \lambda_0}$ and $d_1$ choices.
I.e., the number of queues with $\geq i$ $P_1$-jobs converges to $$\left (\frac{\lambda_1}{1- \lambda_0}\right)^{\frac{d_1^i-1}{d_1-1}}.$$

\end{enumerate}
\end{theorem}

\begin{proof}
Let $q_{0,i}(t)$ and $q_{1,i}(t)$ be the number of $P_0$ jobs and $P_1$ jobs in queue $i$ at time $t$. Let $Q_0(t)=(q_{0,1}(t),\dots,q_{0,n}(t))$ and $Q_1 (t)=(q_{1,1}(t),\dots,q_{1,n}(t))$.
Given that the scheduler only uses current information, the pair $Q(t)=(Q_0(t),Q_1 (t))$ defines a continuous time, discrete space Markov chain.

Our analysis consists of two parts. We first show that the system is positive recurrent with a unique invariant probability measure.  We then formulate and solve a recurrence relation that characterizes the steady state behavior of the system.

Following \cite{AKU24} our proof uses the Lyapunov drift criterion 
for the exponential ergodicity of a continuous 
time Markov process developed by Meyen and Tweedie~~\cite{meyn2012markov,dmt-euemp-95,Meyn_Tweedie_1993}. 

To apply their method we define the \emph{generator} $\mathcal{A}$ of the process $\Phi$ 
as a linear operator on functions $F:\Sigma\to{\bf R}$
defined by
\[\mathcal{A}F(x)=
\lim_{h\downarrow0}\frac{E(F(\Phi(h))\,|\,\Phi(0)=x)-F(x)}h\]
whenever the above limit exists for all $x\in\Sigma.$

To complete the proof of Theorem~\ref{thm:priorities} we invoke the following theorem as a technical tool; it
follows from the more general results
in~\cite{dmt-euemp-95}, specialized to the 
case of a continuous-time Markov process with a countable 
state space.
\begin{theorem}\label{th:drift}
{\em \cite{dmt-euemp-95} }
Consider a Markov process
evolving on a countable
state space that is nonexplosive~\footnote{Recall that a process is explosive if it generates, with probability 1, infinitely many jumps in a finite time interval.},
irreducible (with respect to 
the counting measure on $\Sigma$)
and aperiodic.
If there exists a finite set $C\subset\Sigma$,
constants $b<\infty$, $\beta>0$ and a 
function $V:\Sigma\to[1,\infty)$, 
such that,
\begin{equation}
\label{eq:drift}
\mathcal{A}V(x)\le-\beta
V(x)+b\mathbf{1}_C(x)\qquad{x\in\Sigma}\,,
\end{equation}
then the process is positive recurrent with 
some invariant probability measure $\pi$, 
and there exist constants $\gamma<1$, $D<\infty$ 
such that 
\[\sup_y\lvert P^t(x,y)-\pi(y)\rvert\le D\, V(x)\gamma^t,
\;\;\mbox{for all $t\geq 0$ and all $x\in\Sigma$.}
\]
\end{theorem}

Our system is clearly nonexplosive,
irreducible, and aperiodic. To prove convergence we define a drift function function
$$V(Q (t)) = \sum_{i=1}^n \beta^{q_{0,i}(t) +q_{1,i} (t)},$$ for a constant $\beta= \frac{2}{1+\lambda}>1$.
Let $\lambda= \lambda_0+\lambda_1$, then
\begin{eqnarray}
\mathcal{A}V(Q (t)) &\leq &\frac{1}{n} (\beta^{-1} -1)\sum_{i=1}^n \beta^{q_{0,i}(t) +q_{1,i} (t)} +  (1-\beta^{-1})+\frac{ \lambda}{n}(\beta-1)\sum_{i=1}^n \beta^{q_{0,i}(t) +q_{1,i} (t)}\label{AV-h}\\
&\leq& \frac{1}{n}(\beta^{-1} -1+ \lambda(\beta-1))\sum_{i=1}^n \beta^ {q_{0,i}(t) (i)+q_{1,i}(t)} + \lambda (1-\beta^{-1}) \nonumber\\
&\leq& \frac{1}{n}(-(1+\lambda) + \beta^{-1} +\lambda\beta) V(Q(t)) + (1-\beta^{-1}) \nonumber
\end{eqnarray}
The second term in the right side of equation (\ref{AV-h}) corrects the first term for queues with 0 jobs at time $t$. In the third term we used the fact that the change in adding a new job is dominated by placing it uniformly at random.  
Applying $\beta =\frac{2}{\lambda+1}$ we have 
\begin{eqnarray*}
\mathcal{A}V(Q (t)) &\leq &
-\frac{1}{2n(1+\lambda)}((1+\lambda)^2 -4\lambda)V(Q(t))+\lambda (1-\frac{\lambda+1}{2})\\
&=& -\frac{(1-\lambda)^2}{2n(1+\lambda)}V(Q(t))+\lambda (1-\frac{\lambda+1}{2})
\end{eqnarray*}

We now observe that for $m=\sum_{i=1}^n (q_{0,i}+q_{1,i}(t))$,
$$V(Q (t)) =\sum_{i=1}^n \beta^{q_{0,i}+q_{1,i}(t)}\geq \beta^{\frac{\sum_{i=1}^n (q_{0,i}+q_{1,i}(t))}{n}}= \beta^{\frac{m}{n}}.$$
Therefore, for sufficiently large constant $c$, if the total number of jobs in the system satisfies $m>cn\log n$,
then 
$$\frac{(1-\lambda)^2}{2n(1+\lambda)}V(Q(t))-\lambda (1-\frac{\lambda+1}{2})\geq \frac{2(1-\lambda)^2}{2n(1+\lambda)}V(Q(t))$$

and 
$$\mathcal{A}V(Q (t)) \leq -\frac{2(1-\lambda)^2}{2n(1+\lambda)}V(Q(t))+ {\bf 1}_{\{Q~|~m\leq cn\log n\}} \lambda(1-\beta^{-1}),$$
satisfying the conditions of Theorem~\ref{th:drift}.



Next we analyze the stationary distribution of the process. 
Denote the fraction of queues with at least $i$ $P_0$-jobs at time $t$
$$H_{\geq i} (t)= \frac{1}{n}|\{ j\in[1,n]~|~q_{0,j}(t)\geq i \}|,$$ 
and
$$H_{0} (t)= \frac{1}{n}|\{ j\in[1,n] ~|~q_{0,i}=0\}|.$$ 
Similarly, 
denote the fraction of queues with at least $i$ $P_1$-jobs at time $t$
$$L_{\geq i} (t)= \frac{1}{n}|\{\{ j\in[1,n]~|~q_{1,j}(t)\geq i \}|,$$
and
$$L_{0} (t)= \frac{1}{n}|\{  j\in[1,n] ~|~q_{1,i}=0\}|,$$
Since the execution of $P_0$-jobs are not delayed by $P_1$-jobs, we know that in the stationary distribution 
$$E[H_{\geq i} ] =\lambda_0^{\frac{d_0^j -1}{d_0-1}}~~~\mbox{for} ~~~i\geq 1,~~~~~
\mbox{and}~~~~
E[H_0]= 1-\lambda_0.$$

Since $P_1$-jobs are executed only in queues with no $P_0$-jobs, we focus on the fraction of queues with no $P_0$-jobs and $\geq k$ $P_1$-jobs, which we denote 

$$L_{0,\geq k} (t)= \frac{1}{n}|\{ j\in[1,n]~|~q_{0,j}(t)= 0 ~\mbox{and}~ q_{1,j}(t)\geq k\}|,$$

With these notations we have: 
$$\sum_{k\geq 0} (E[ L_{0,\geq k} ]-E[L_{0,\geq k+1} ])=E[L_{0,\geq 0} ]= E[H_0] = 1-\lambda_0$$
and for all $k\geq 1$,
$$\lambda_1 (E[L_{\geq k-1} ]^{d_1}-E[L_{\geq k} ]^{d_1}) =  (E[L_{0,\geq k}] -E[L_{0,\geq k+1}] ),$$
or 
$$E[L_{0,\geq k}]= \lambda_0 E[L_{\geq k-1} ]^{d_1}.$$

It's not hard to verify that 
$$E[L_{\geq k-1} ]^{d_1}=E[L_{0,\geq k}] = (1-\lambda_0)E[L_{\geq k}]$$
satisfies these relations. Thus,
$$E[L_{\geq k}] = \left (\frac{\lambda_1}{1- \lambda_0}\right)^{\frac{d_1^{k}-1}{d_1 -1}}, ~~~~\mbox{for}~k\geq 1.$$
\end{proof}
Similar, but more tedious analysis gives the result for three priorities:

\begin{theorem}
\label{th:3-priorities}
Suppose that we have jobs with three priorities, arriving with rates $\lambda_0$, $\lambda_1$, and $\lambda_2$. 
\begin{enumerate}
\item 
The distribution of the number of highest priority jobs in queues  follows the balanced allocation process with $\lambda_0$ and $d_0$. I.e., the number of queues with $\geq i$ $P_0$-jobs converges to $$\lambda_0^{\frac{d_0^i-1}{d_0-1}}.$$
\item
The distribution of the number of jobs with priority 1 in queues  follows the balanced allocation process with $\frac{\lambda_1}{1- \lambda_0}$ and $d_1$.
I.e., the number of queues with $\geq i$ $P_1$-jobs converges to $$\left (\frac{\lambda_1}{1- \lambda_0}\right)^{\frac{d_1^i-1}{d_1-1}}.$$

\item
The distribution of the number of jobs with priority 2 in queues  follows the balanced allocation process with $\frac{\lambda_2}{1- \lambda_0-\lambda_1}$ and $d_2$.
I.e., the number of queues with $\geq i$ 2-jobs converges to $$\left (\frac{\lambda_2}{1- \lambda_0-\lambda_1}\right)^{\frac{d_2^i-1}{d_2-1}}.$$
\end{enumerate}
\end{theorem}

Note that the base of the exponent for the distribution of queues with lower priorities is smaller, and therefore, the double exponential decay is less significant.

\paragraph{Significance:} The analytical results provide a rigorous mathematical validation for the trends observed in the priority simulations results. Specifically, Theorem ~\ref{th:3-priorities} confirms the simulation findings that $P_0$ jobs operate as if in an isolated system, unaffected by lower-priority traffic, while $P_1$ and $P_2$ jobs still benefit from the double exponential decay characteristic of $d$-way balanced allocation. The asymptotic values calculated in Table~\ref{tab:computed_priority} show that for $d=2$, the expected queue length for the lowest priority ($P_2$) is approximately 1.99, which aligns with the simulated average depth of 2.11 to 2.54 across the various priority strategies. This close correspondence between the analysis and simulations demonstrates that the "Independent" and "MineThenTotal" strategies effectively distribute lower-priority load without the need for complex job migrations~\cite{9812880}, ensuring that the system remains mathematically stable and efficient even for background tasks.
\begin{table}[h!]
\centering
\begin{subtable}[t]{0.45\textwidth} 
\centering
\begin{tabular}{c|ccc}
\hline
$d$ & $P0$ & $P1$ & $P2$ \\
\hline
$1$ & $0.46341$ & $0.86364$ & $6.33333$ \\
$2$ & $0.34874$ & $0.56753$ & $1.98778$ \\
$3$ & $0.32672$ & $0.50958$ & $1.57149$ \\
$4$ & $0.31985$ & $0.48479$ & $1.39012$ \\
\hline
\end{tabular}
\caption{Expected queue length}
\end{subtable}
\hfill
    \begin{subtable}[t]{0.45\textwidth}
\centering
\begin{tabular}{c|ccc}
\hline
$d$ & $P0$ & $P1$ & $P2$ \\
\hline
$1$ & $7$ & $11$ & $65$ \\
$2$ & $3$ & $4$ & $7$ \\
$3$ & $3$ & $3$ & $5$ \\
$4$ & $3$ & $3$ & $5$ \\
\hline
\end{tabular}
\caption{Maximum queue length}
\end{subtable}
\caption{Asymptotic values computing from Theorem~\ref{th:3-priorities} for $d= d_0=d_1=d_2$ and $\lambda_0=\lambda_1=\lambda_2=0.95/3$.}
\label{tab:computed_priority}
\end{table}

\section{Noisy information on queue sizes}
\label{sec:fuzz}
So far our simulations and analysis assume perfect knowledge about the number of jobs in each queue.  In production environments this is generally not the case.  For example, if we must schedule across globally distributed servers, fetching the counts of $d$ queues may be prohibitive for both cost and latency (e.g., it may take ~30+ms for one server to reach another server in the same continent).  Thus, production systems must operate with imperfect information.  For example, rather than explicitly getting counts on every request, a scheduler may get the count information when it actually sends a request to a server and uses that, possibly stale, information when deciding whether to send another request to that server.  This section models imperfect information into our simulations and analysis.  

We explore two models of imperfect information in this section: \textit{Lag} and \textit{Fuzz}.
\textit{Lag} means that we use older snapshots of the states of queues when picking between queues, rather than getting the current state.  This explicitly models the situation where it is prohibitive to get the fresh state of the system for every request.  \textit{Fuzz} corrupts comparisons: when comparing queue depths, two values that are within the fuzz are considered the same. 

\subsection{Simulations}
To simulate \textit{Lag} our simulation takes a snapshot every 2000 jumps and at each jump it uses the appropriate snapshot based on the desired lag.  For example, when simulating a lag of $10,000$, our simulation uses the snapshot at $20,000$ for jumps from  $30,000$ to $31,999$ and then uses snapshot at $22,000$ for the next 2000 jumps, and so on.  This is analogous to periodic updates in prior work~\cite{mitzenmacher2000useful}.
 We explore fuzz values of $2$ and $10$.

\begin{table}[h!]
\centering
\begin{tabular}{|c|r|r|r|r|r|}
\hline
$d$ & Baseline & Lag=2000 & Lag=10000 & Fuzz=2 & Fuzz=10 \\
\midrule
1 & 18.75 & 19.11 & 18.87 & 19.12 & 19.15 \\
2 & 3.24 & 3.84 & 5.88 & 3.98 & 7.16 \\
3 & 2.40 & 3.40 & 6.57 & 3.26 & 6.57 \\
4 & 2.11 & 3.40 & 7.42 & 2.93 & 6.27 \\
\hline
\end{tabular}
\caption{Average Queue Depth (No Bursts, 1 Priority). Comparison of Baseline, Fuzz, and Lag.}
\label{tab:avg_fuzz_lag}
\end{table}

\begin{table}[h!]
\centering
\begin{tabular}{|c|r|r|r|r|r|}
\hline
$d$ & Baseline & Lag=2000 & Lag=10000 & Fuzz=2 & Fuzz=10 \\
\midrule
1 & 136.30 & 154.07 & 143.53 & 145.13 & 150.53 \\
2 & 6.77 & 9.67 & 19.23 & 9.53 & 19.93 \\
3 & 4.70 & 9.90 & 22.53 & 7.47 & 17.23 \\
4 & 4.00 & 10.90 & 26.33 & 6.80 & 16.00 \\
\hline
\end{tabular}
\caption{Max Queue Depth (No Bursts, 1 Priority). Comparison of Baseline, Fuzz, and Lag.}
\label{tab:max_fuzz_lag}
\end{table}

Tables ~\ref{tab:avg_fuzz_lag} and \ref{tab:max_fuzz_lag} explores the performance of the two error models.  "Baseline" is without any lag or fuzz and is the same configuration as Table~\ref{tab:simplest}; it is there to provide a point of comparison. We see that both lag and fuzz  degrade all $d$, but $d>1$ still has much lower average and maximum queue depths compared to $d=1$.\footnote{We see some variation in the row for $d=1$ even though this case is unaffected by lag or fuzz; this variation is statistical noise despite doing 30 runs.}  A lag of 10000 or fuzz of 10 are the worst in this regard.  A lag of 2000 jumps means that that we have had about 1000 jumps of arrivals and about 1000 of departures since our last snapshot.  Since $\lambda=0.95$, our laggy data is 950 arrivals and 950 departures behind the simulation.  In a system of 10000 queues and balanced scheduling, the queue count from a snapshot may be off from the actual count by ~$\pm1$.  Since fuzz of 2 considers queues with depths that are within 2 to be the same, it is not surprising that fuzz of 2 performs comparably to lag of 2,000.

We had expected that a high level of fuzz $d>1$ will perform as bad as Random since increasing fuzz forces more and more of the choices to be random.  Surprisingly we do not see that with fuzz of 10: $d>1$ still outperforms $d=1$; we also explored (but do not include in the table) a fuzz of 20 and even then $d>1$ performs better than Random.  This means that even with a high value of fuzz, $d>1$ is able to avoid the really deep queues and that makes it perform better than $d=1$.

\begin{figure}[htb]
    \centering
    \includegraphics[width=0.6\textwidth]{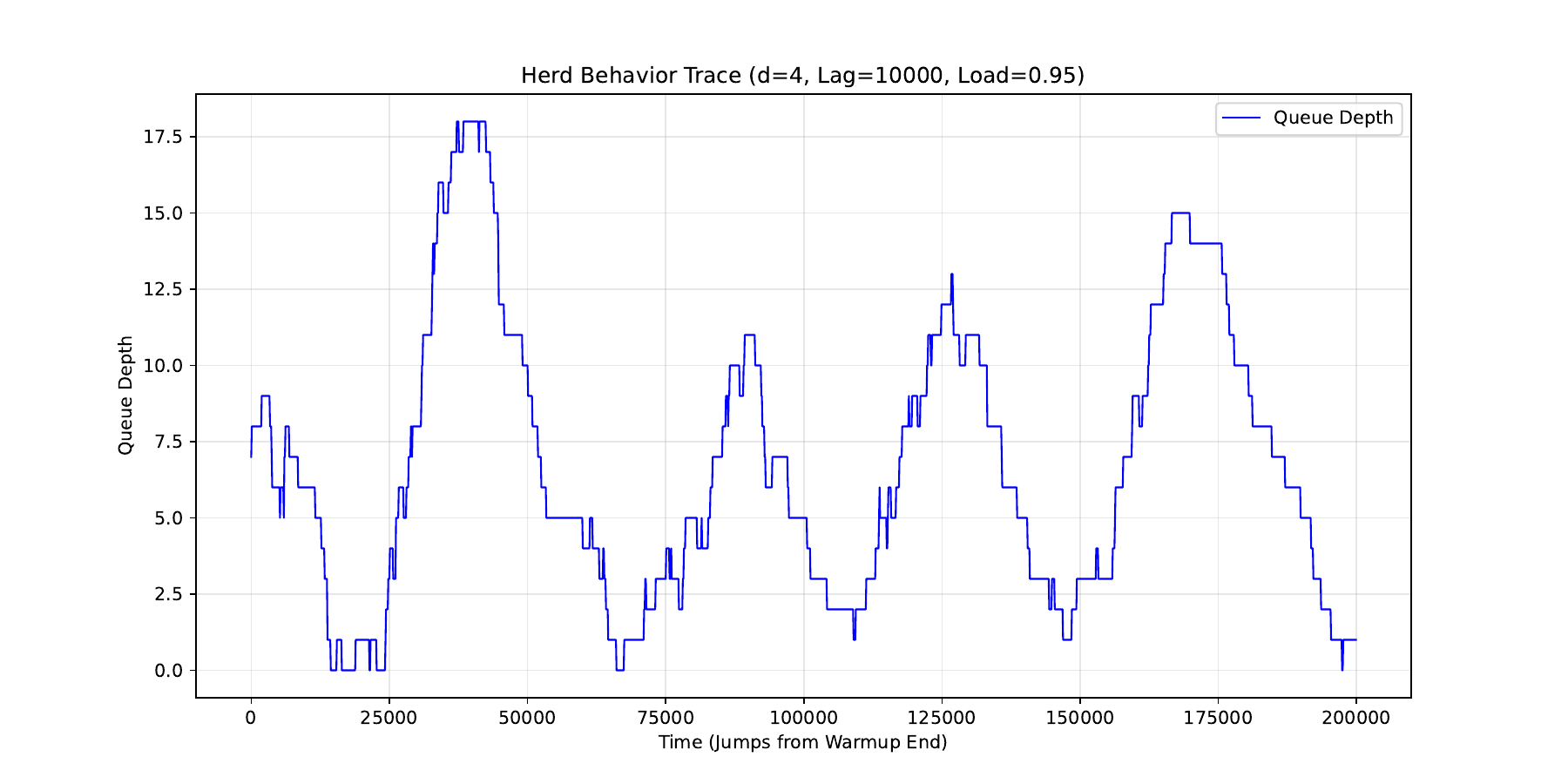}
    \caption{Queue depth for one sample queue with $d=4$ (no bursts, 1 priority, lag=10000).} 
    \label{fig:lag_anomaly_d4}
\end{figure}

A lag of 10000 exposes the "herd behavior" anomaly~\cite{dahlin2000interpreting,mitzenmacher2000useful}: as we increase $d$ the average queue depth grows  because the laggy information means that \powerOfD\ will send many requests to formerly empty queues, and not realize that it has overloaded them before the lag period has passed.  Figure~\ref{fig:lag_anomaly_d4} demonstrate this phenomenon for a sample queue.  For readability we only show part of the simulation.  The depth of the queue exhibits sawtooth behavior: it accumulates a significant depth, then it drains to empty or nearly empty, and then the cycle starts again.  We can see that the accumulation is quick (steep increase in queue depth) but the drain slope is less steep.  This is because the drain rate depends on the total number of queues and $\lambda$, rather than number of jobs in the particular queue.  Finally a higher $d$ makes it more likely that the scheduler will find empty queues and load them disproportionately.

\begin{figure}[htb]
    \centering
    \includegraphics[width=\textwidth]{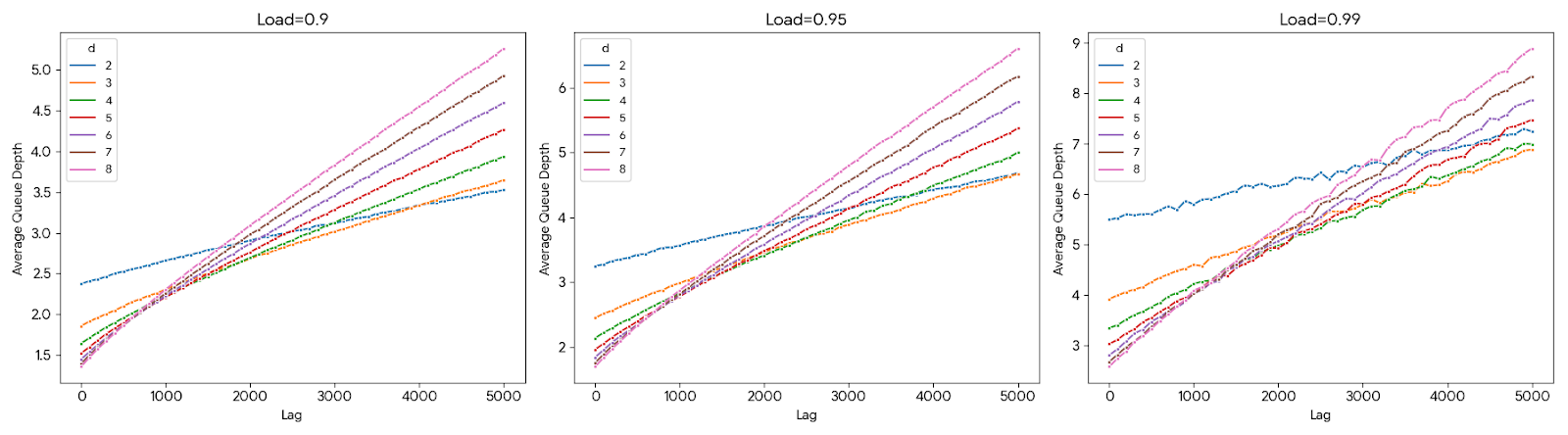}
    \caption{$d$ versus lag and $\lambda$} 
    \label{fig:d_v_lag}
\end{figure}

Figure~\ref{fig:d_v_lag} extends this analysis with a broader exploration of the space.  Each plot represents a particular load ($\lambda$) and each curve is for a single $d$.  A point $(x,y)$ on the curve says that at a lag of $x$, the average queue depth is $y$.  We omit $d=1$ from these curves since it is not affected by the lag and its curve is consistently high and flat compared to the curves in this plot.

We see two trends in these graphs.  First, for small values of lag higher $d$ and in particular $d>3$ significantly outperform $d=2$ or $d=3$.  The crossover when this stops being the case depends on the load: the higher the arrival rate ($\lambda$) the higher lag we need before low values of $d$ outperform higher values of $d$.  Second, the curves for higher $d$ have higher slopes.  This cuts both ways: (i) as lag increases, higher $d$ degrade more rapidly; but (ii) as lag decreases, higher $d$ significantly outperforms lower $d$.  Thus, one should pick the $d$ carefully based on the expected load of the system and the amount of staleness that we have engineered into the system.

\subsection{Analysis}



 We show that for any constant fuzz $b\geq 1$, the tail distribution of the steady state lengths of the queues is still governed by a double exponential decaying function. Because the constant $\beta>1$ depends on $b$ and $d$ the expression for ``large'' $s_i$ is a bit difficult to interpret; we explicate this dependency in Table~\ref{tb:beta}.

\begin{theorem}\label{th:fuzz}
Consider a system with $n$ queues, Poisson arrival rate $\lambda<1$, exponential service time with expectation 1, running a balanced allocation schedule with $d=2$, and fuzz error $b$. Let $s_i$ be the fraction of queues with at least $i$ jobs. In the steady state, 
\begin{enumerate}
    \item $s_i =  \lambda^i$, for $i\leq b+1$ 
    \item $s_i\approx 
    \lambda^{(b+1+\frac{1}{d-1})\beta^{i-b}-\frac{1}{d-1}}$
    for $i>b+1$, and some constants  $\beta >1$.
\end{enumerate}
The constant $\beta$ is a function of $b$ and $d$, but does not depend on $i$ or $n$.
\end{theorem}
\begin{proof}
%
We formulate a differential equation that represents the dynamic (in expectation) of the system in steady state:
\begin{eqnarray*}
     \frac{ds_i}{dt} &\approx& 
     \lambda(s_{i-1}-s_{i})s^{d-1}_{i-1-b}-(s_i-s_{i+1}) =0, 
~~~~~~~\mbox{for}~i> 1 \\
     s_1 &=& \lambda\\
     s_0 &=& 1. 
 \end{eqnarray*}

The first equation represents the fact that in a steady state the expected change in $s_i$ is 0. An increase in $s_i$ requires that (1) all $d$ probes choose queues with at least $i-b-1$ jobs, (2) that not all probes choose queues with more than $i+b-1$ jobs, and (3) that a random choice among the probes that choose queues with 
$i+b-1>\ell>i-b-1$ jobs is a probe that choose a queue with $i-1$ jobs.

Thus, the expected increase in $s_i$ is 
$$s^+_i = \lambda (s_{i-b-1}^d-s_{i+b-1}^d)\frac{s_{i-1}-s_i}{s_{i-b-1}-s_{i+b-1}}\approx \lambda(s_{i-1}-s_{i})s^{d-1}_{i-1-b}$$
which must equal the expected decrease, $s_{i}-s_{i+1}$. 

The boundary conditions reflect the fact that no queue has a negative number of jobs, and the number of non-empty jobs (the expected number of departing jobs) must equal the arrival rate $\lambda$.


Since the distribution of the $s_i$'s is dominated by the distributions of queue length in random assignment, $s_i$ decreases at least exponentially in $i$. We therefore further approximate
$s_i-s_{i+1}\approx s_i$,
giving the recurrence relation 
$$s_i \approx \lambda s_{i-1}s^{d-1}_{i-1-b}.$$

For $1\leq i\leq b+1$, $s^{d-1}_{i-1-b}=1$ and 
$$s_i\approx \lambda^i.$$

For $i>b+1$,
taking the logarithm base $\lambda$, we get
\begin{equation}
     \log(s_{i}) - 1-\log(s_{i-1}) -(d-1) \log (s_{i-b-1})  = 0
\end{equation}
A standard technique for linear homogeneous recurrence with constant coefficients~\cite{rosen2019discrete} gives 
$$\log(s_{i}) \approx a \cdot \beta^i +c,$$
where $\beta>1$ is the unique positive root of the polynomial $r^{b+1}- r^{b}-(d-1)=0$ (Descartes' Rule of Signs), $a$ is determined by the boundary conditions and $c$ is the particular constant solution.

Thus, $$s_i \approx e^{a\beta^i +c} .$$

Applying the boundary condition $s_{b+1} =\lambda^{b+1}$ we compute $a=(b+1)\log \lambda/\beta^{b+1}<1$. The particular solution is $-\log \lambda/(d-1)$. Thus, for $i>b$,  
$$s_i \approx \lambda^{(b+1+\frac{1}{d-1})\beta^{i-b-1}-\frac{1}{d-1}}. $$
\end{proof}

To see that $\beta$ tends to 1 for large $b$, note that $\beta$ is a root of $
r^{b}=1/(r-1).$ As $b$ grows, the value of $r$ must get close to 1. Table \ref{tb:beta} gives the values of $\beta$ for various values of $b$ and $d$.

\begin{table}[h!]
\centering
\begin{tabular}{|c|c|c|c|}\hline\textbf{$fuzz=b$} & \textbf{$d$} & \textbf{$\beta$} & \textbf{Asymptotic expected queue size} \\ \hline  
1  & 2 & 1.618  & 3.25512 \\ \hline 
2  & 2 & 1.466  & 4.15549 \\ \hline 
10 & 2 & 1.184  & 8.79702 \\ \hline 
1  & 3 & 2.000  & 1.58037 \\ \hline 
2  & 3 & 1.696  & 2.57666 \\ \hline 
10 & 3 & 1.237  & 7.59614 \\ \hline 
1  & 4 & 2.302  & 0.94372 \\ \hline 
2  & 4 & 1.863  & 1.94949 \\ \hline 
10 & 4 & 1.2715 & 7.05114 \\ \hline
\end{tabular}
\caption{The principal root $\beta$ and the asymptotic expected queue length for various values of $b$ and $d$. Computed with MatLab R2025b based on Theorem~\ref{th:fuzz}.}
\label{tb:beta}
\end{table}

\paragraph{Significance:} Rigorous analysis of the effect of lag is challenging because the probability of a change in an individual queue at a given jump and lag is hard to quantify. However, our simulations show that performance with fuzz errors could serve as a reasonable approximation, while ignoring the secondary effect of larger $d$ values. Our analysis  characterizes the effect of fuzz error. It proves that the distribution of queue length is well approximated by a convolution of two distributions. Queues with depth smaller than the fuzz follow an M/M/1 queue distribution (exponential decay). Beyond that threshold, queue depth follows a double exponential decay distribution. The rate of decay is a function of the fuzz, and is slower for larger fuzz values.

\section{Putting it all together}
\label{sec:all}
So far we have studied each variant in isolation: the number of bursts, multiple priorities, and tolerance to imperfect information.  This section shows that \powerOfD{} is significantly better than Random even if we superpose these variants.

\begin{figure}[htb]
    \centering
    \begin{minipage}{0.48\textwidth}
        \centering
        (a) No lag 
        \includegraphics[width=\textwidth]{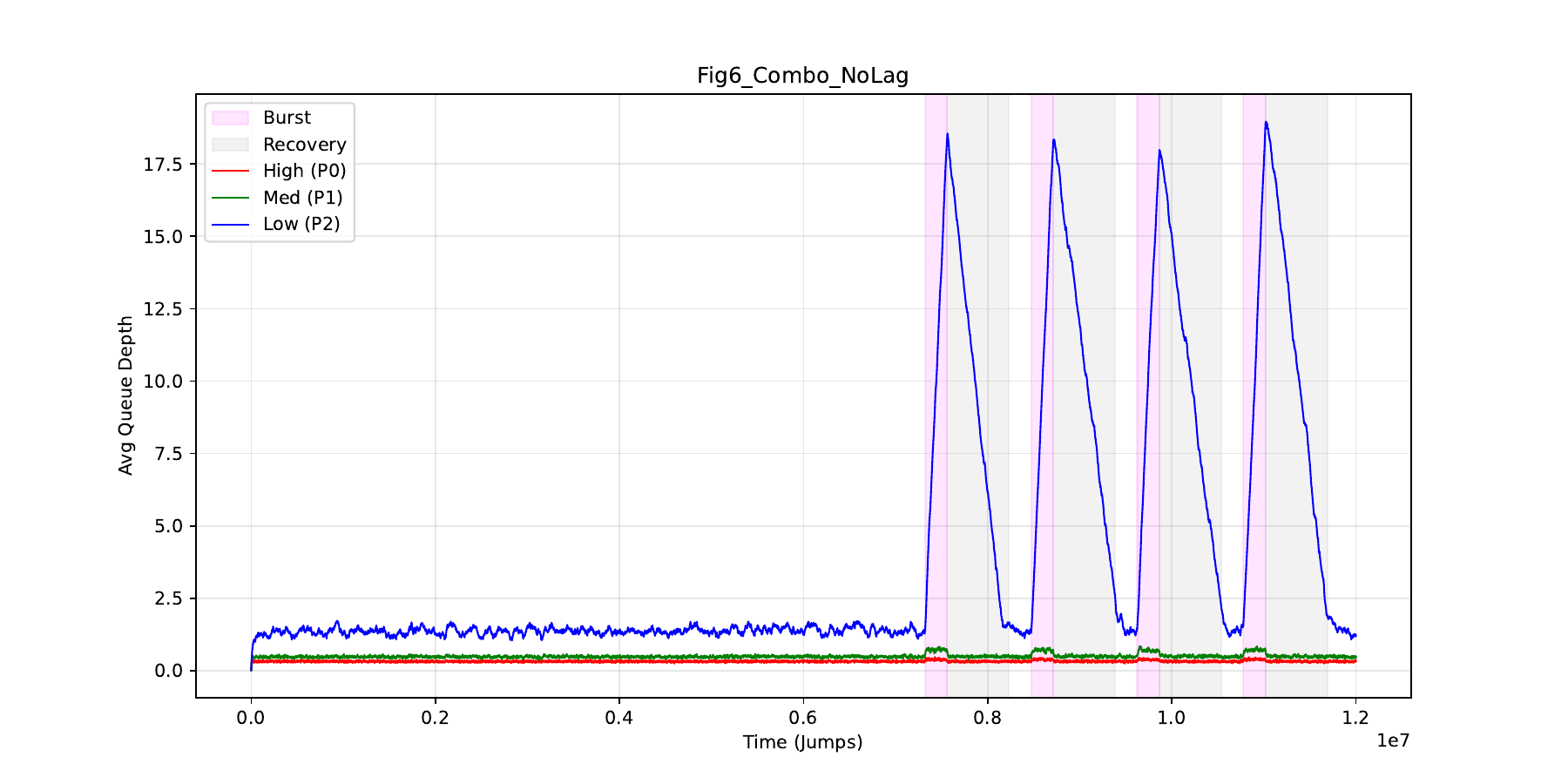}
    \end{minipage}\hfill
    \begin{minipage}{0.48\textwidth}
        \centering
        (b) Lag=2000. 
        \includegraphics[width=\textwidth]{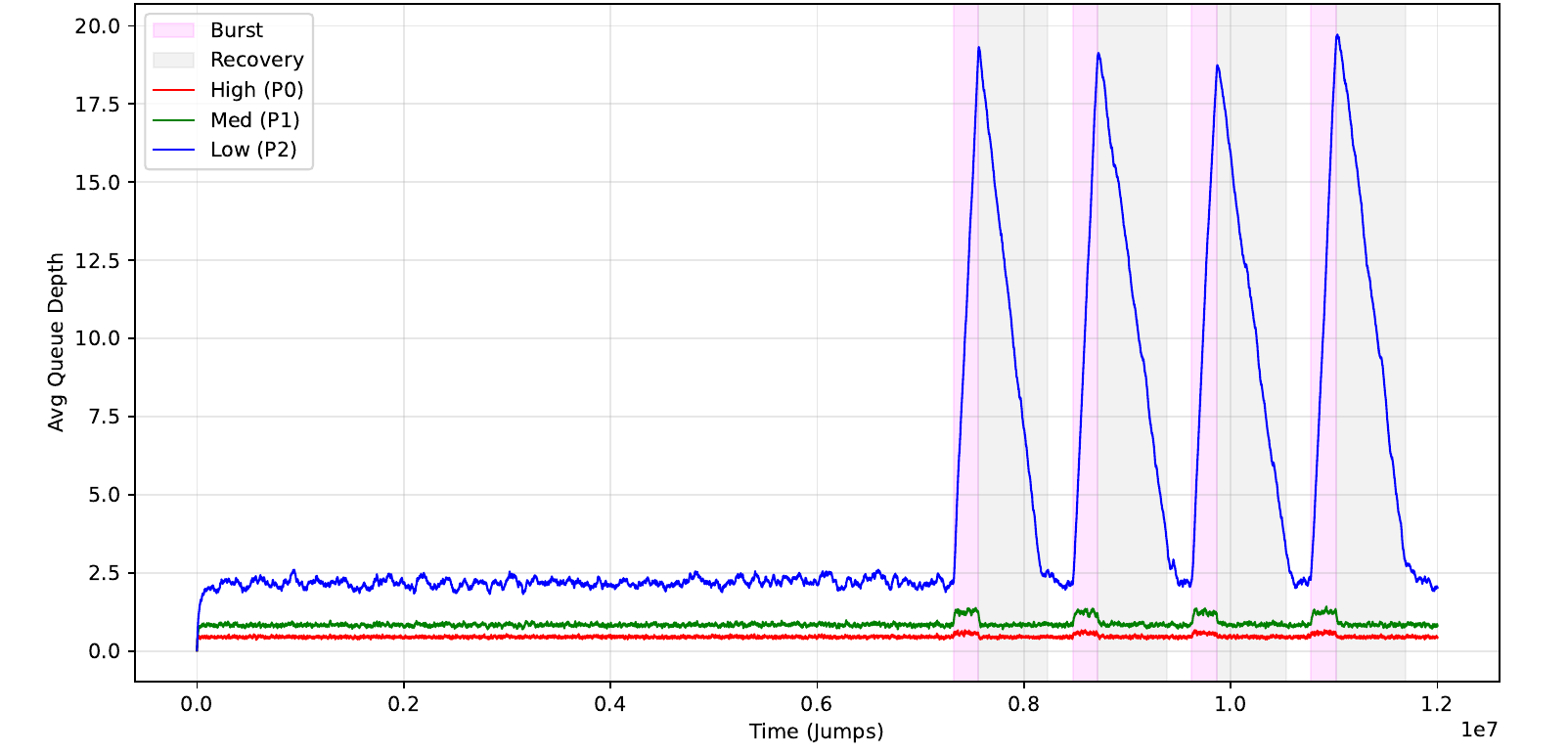}
    \end{minipage}
    \vspace{-5pt}
    \caption{Average queue depth. 4 bursts, 3 priorities, MineThenTotal} 
    \label{fig:aggregate}

\end{figure}

To keep this exploration manageable, we pick one reasonable setting for each variable: $\lambda=0.95$, $d=4$, with 4 bursts, 3 priorities with MineThenTotal, and lag=$2000$ (recall that lag=$2000$ behaves similarly to fuzz$=2$).  Figure~\ref{fig:aggregate}(a) and Table~\ref{tab:all-together}(a) show this data without any lag and Figure~\ref{fig:aggregate}(b) and Table~\ref{tab:all-together}(b) show this data with a lag=$2000$.

Figure~\ref{fig:aggregate}(b) shows that even with lag, $d=4$  retains a low number of queued items and the quick recovery from bursts at all priorities.  Interestingly, higher priorities (all except for $P2$) barely suffer during bursts; $P2$ takes the brunt of the impact.  Comparing Figure~\ref{fig:aggregate}(a) to Figure~\ref{fig:aggregate}(b) we see that adding lag shifts up all curves.

\begin{table}[htb]
    \centering
    \footnotesize 
    \begin{minipage}{0.48\textwidth}
        \centering
        (a) Average
        \begin{tabular}{l|rrrrrr}
            \toprule
            & \multicolumn{3}{c}{Lag=0} & \multicolumn{3}{c}{Lag=2000} \\
            \cmidrule(lr){2-4} \cmidrule(lr){5-7}
            $d$ & P0 & P1 & P2 & P0 & P1 & P2 \\
            \midrule
            1 & 0.47 & 1.27 & 30.64 & 0.46 & 1.27 & 30.61 \\
            2 & 0.35 & 0.65 & 2.54 & 0.41 & 0.81 & 2.93 \\
            3 & 0.33 & 0.54 & 1.69 & 0.43 & 0.79 & 2.31 \\
            4 & 0.33 & 0.49 & 1.41 & 0.45 & 0.84 & 2.20 \\
            \bottomrule
        \end{tabular} \\
    \end{minipage}\hfill
    \begin{minipage}{0.48\textwidth}
        \centering
        (b) Maximum
        \begin{tabular}{l|cccccc}
            \toprule
            & \multicolumn{3}{c}{Lag=0} & \multicolumn{3}{c}{Lag=2000} \\
            \cmidrule(lr){2-4} \cmidrule(lr){5-7}
            $d$ & P0 & P1 & P2 & P0 & P1 & P2 \\
            \midrule
            1 & 7.03 & 16.27 & 218.13 & 7.23 & 16.93 & 213.10 \\
            2 & 2.77 & 3.03 & 6.13 & 4.60 & 5.10 & 7.83 \\
            3 & 2.13 & 2.63 & 4.00 & 5.13 & 5.73 & 6.93 \\
            4 & 2.13 & 2.33 & 3.30 & 5.67 & 6.10 & 7.17 \\
            \bottomrule
        \end{tabular}\\
    \end{minipage}
    \vspace{10pt}
    \caption{Queue Depth (4 Bursts, MineThenTotal Strategy)         \label{tab:all-together}
}

\end{table}

\section{Related work} \label{sec:related}

\subsection{Foundational Work on $d$-way Balanced Allocation}
The combinatorial balanced allocation process was first formalized and analyzed in~\cite{ABKU}. This paper proved the double exponential decay phenomenon for the static problem (no deletions) and the dynamic version with random deletions. A queuing version of the problem where jobs arrive according to a Poisson process and have exponential service time was studied by 
Vvedenskaya et al.~\cite{vvedenskaya1996queueing} and Mitzenmacher~\cite{963420}, who utilized differential equations to characterize an invariant distribution where the fraction of queues with at least $i$ jobs decays doubly exponentially.

\subsection{Heavy Loads and Burst Recovery}
Our analysis of burst recovery builds upon the ``heavily loaded'' case analyzed by Berenbrink et al.~\cite{10.1145/335305.335411}. They demonstrated that for static allocation of $m$ balls into $n$ bins, the maximum load differs from the average by only $O(\log \log n)$, independent of $m$. Our work extends this to burst recovery in a dynamic setting. Specifically, we leverage recent advances by Berenbrink et al.~\cite{berenbrink2025ballsbinsinfiniteprocess} regarding infinite processes with random deletions to bound the recovery time of queues following a burst event. While prior work focused on the steady-state behavior under heavy load, we explicitly model the transient \textit{recovery} phase where the arrival rate drops from $\lambda_{high}$ back to $\lambda_{low}$.

\subsection{Priority Scheduling}
Incorporating priorities into randomized load balancing has gained attention recently. 
Mitzenmacher et al.~\cite{Mitzenmacher2001Power} demonstrated the flexibility of the fluid limit approach by formulating a differential queuing system for a two priorities system, where $2$-way balanced allocation is used for the high priority jobs and random scheduling is used or the low priority jobs. Niu et al~\cite{9812880} analyzed a system with two priorities, where low priority jobs are placed using $2$-way balanced allocation that considered the total load of the queues, while high priority jobs are placed using $d$-way balanced allocation that considered only high priority jobs. However in their model, when a high priority job is not placed in a queue with minimum total load among the $d$ random choices, one low priority job is removed from that queue and moved to a server with lowest total load among the $d$ choices. Moving jobs that are already in a queue is an expensive operation. We show that similar performance is obtained without this extra assumption. A different application of $d$-way balanced allocation was studied in~\cite{10.1145/3087801.3087810} where $d$ queues are sampled to schedule a high priority job from multiple concurrent priority queues. Alistarh et al.~\cite{10.1145/3087801.3087810} explored the ``power of choice'' in priority scheduling, sampling $d$ queues to schedule high-priority jobs. 

\subsection{Imperfect information and herd Behavior}
The use of stale information in the dynamic supermarket model was studied by Mitzenmacher~\cite{mitzenmacher2000useful} and Dahlin~\cite{dahlin2000interpreting}. Mitzenmacher demonstrating that under high delay, increasing $d$ can actually degrade performance and proposes mitigation strategies. Dahlin further examined average response time in a single burst configuration.  Our work builds upon this prior work by systematically exploring the interaction of lag (staleness) and fuzz (comparison error) specifically in the context of \textit{burst recovery} and \textit{multi-priority} environments. We show that even with significant lag, $d$-way allocation ($d>1$) maintains a substantial advantage over random assignment for burst recovery, provided the lag is not excessive relative to the system load.

\subsection{Other directions}
Join-Idle queues~\cite{lu2011jiq} is an alternative to \powerOfD{} in that it puts the burden on the servers to notify the dispatchers of their idleness.  It  requires two load balancing layers: one to to pick a dispatcher when a new job arrives and one to pick a dispatcher to notify when a processor becomes idle. The benefit of this approach is that it avoids the cost of getting the queue depth of $d$ server when a job arrives and instead this work is effectively asynchronous when a job completes.  As with our work, they use a combination of simulations and analytical analyses but do not extend their evaluation with priorities, bursts, or noisy information.

\section{Conclusions}
Mathematically, \powerOfD{} is characterized by a double exponential decay in the probability that a queue has at least $i$ jobs. A major contribution of our work is proving that this still holds in a variety of settings (bursts, priorities, and noise) arising in large-scale cloud services. Our analysis in Section~\ref{sec:bursts} gives asymptotic bounds showing a separation in queue depths between the cases $d=1$ and $d>1$.
The analyses in Sections~\ref{sec:priorities} and \ref{sec:fuzz} precisely characterize the distributions of queue depths under jobs of varying priorities, and noisy queue depth measurements. In these cases we show close matches between simulated and analytically predicted queue depths.
Under all variants, our simulations highlight the benefits of \powerOfD{}, with a nuanced portrayal of how the choice of $d$ is sensitive to the latency in queue depth information (Section~\ref{sec:fuzz}).

\bibliography{Queues}{}
\bibliographystyle{plain}
\appendix
\section{Appendix}
Most of the data in this paper is for a system that is running hot ($\lambda=0.95)$ and for bursts that push the system above the capacity of the system. We have explored other $\lambda$ values; in this appendix we present simulation results for $\lambda=0.75$ with all other parameters unchanged.  In this case the bursts (which are $1.2 * 0.75 = 0.9$) do not push the system above its capacity. The tables below are analogous to the ones in the main body of the paper and clearly demonstrate that our results are relatively insensitive to the specific load on the system.  


\begin{table}[h!]
\label{tab:baseline}
\begin{tabular}{l|rr}
\toprule
$d$ & Avg Depth & Max Depth \\
\midrule
1 & 3.02 & 24.50 \\
2 & 1.31 & 4.10 \\
3 & 1.11 & 3.00 \\
4 & 0.98 & 2.93 \\
\bottomrule
\end{tabular}
\caption{Queue Depth Statistics by $d$ for $\lambda=0.75$}
\label{tab:simplest0.75}
\end{table}

\begin{table}[h!]
\begin{tabular}{lrrrrrrrr}
\toprule
 & \multicolumn{4}{c}{(a) Average Depth} & \multicolumn{4}{c}{(b) Maximum Depth} \\
\cmidrule(lr){2-5} \cmidrule(lr){6-9}
 & 0 & 2 & 3 & 4 & 0 & 2 & 3 & 4 \\
$d$ &  &  &  &  &  &  &  &  \\
\midrule
1 & 3.02 & 2.98 & 3.00 & 3.00 & 24.50 & 25.60 & 28.80 & 29.40 \\
2 & 1.31 & 1.34 & 1.32 & 1.31 & 4.10 & 4.17 & 4.27 & 4.27 \\
3 & 1.11 & 1.10 & 1.09 & 1.09 & 3.00 & 3.00 & 3.03 & 3.03 \\
4 & 0.98 & 0.99 & 1.00 & 1.00 & 2.93 & 2.80 & 3.00 & 3.00 \\
\bottomrule
\end{tabular}
    \vspace{10pt}
    \caption{Impact of Bursts on Queue Depth with $\lambda=0.75$}
    \label{tab:burst0.75}
\end{table}

\begin{table}[h!]
\begin{tabular}{l|rrrrrrrrrrrr}
\toprule
 & \multicolumn{3}{c}{CumulativeThenTotal} & \multicolumn{3}{c}{Independent} & \multicolumn{3}{c}{MineThenTotal} & \multicolumn{3}{c}{TotalThenMine} \\
\cmidrule(lr){2-4} \cmidrule(lr){5-7} \cmidrule(lr){8-10} \cmidrule(lr){11-13}
$d$ & P0 & P1 & P2 & P0 & P1 & P2 & P0 & P1 & P2 & P0 & P1 & P2 \\
\midrule
1 & \textbf{0.34} & \textbf{0.67} & \textbf{2.01} & \textbf{0.34} & \textbf{0.67} & \textbf{2.02} & \textbf{0.33} & \textbf{0.66} & \textbf{2.01} & \textbf{0.33} & \textbf{0.68} & \textbf{1.98} \\
2 & \textbf{0.27} & \textbf{0.39} & 0.70 & 0.29 & \textbf{0.40} & \textbf{0.63} & \textbf{0.27} & \textbf{0.39} & 0.71 & \textbf{0.28} & \textbf{0.40} & 0.66 \\
3 & \textbf{0.25} & \textbf{0.33} & 0.53 & 0.27 & \textbf{0.34} & \textbf{0.46} & \textbf{0.25} & \textbf{0.34} & 0.52 & \textbf{0.26} & \textbf{0.34} & 0.50 \\
4 & \textbf{0.25} & \textbf{0.30} & 0.44 & 0.27 & 0.33 & \textbf{0.40} & \textbf{0.25} & \textbf{0.31} & 0.44 & \textbf{0.25} & 0.32 & 0.43 \\
\bottomrule
\end{tabular}
\caption{Average Queue Depth by $d$ and Priority with $\lambda=0.75$}
\label{tab:priority_avg}
\end{table}

\begin{table}[h!]
\begin{tabular}{l|rrrrrrrrrrrr}
\toprule
 & \multicolumn{3}{c}{CumulativeThenTotal} & \multicolumn{3}{c}{Independent} & \multicolumn{3}{c}{MineThenTotal} & \multicolumn{3}{c}{TotalThenMine} \\
\cmidrule(lr){2-4} \cmidrule(lr){5-7} \cmidrule(lr){8-10} \cmidrule(lr){11-13}
$d$ & P0 & P1 & P2 & P0 & P1 & P2 & P0 & P1 & P2 & P0 & P1 & P2 \\
\midrule
1 & \textbf{5.20} & \textbf{8.57} & \textbf{21.90} & \textbf{5.03} & 9.70 & \textbf{21.97} & \textbf{5.00} & 9.10 & \textbf{21.73} & \textbf{4.97} & 9.00 & \textbf{22.77} \\
2 & \textbf{2.00} & 2.90 & 3.53 & 3.03 & 3.10 & 3.27 & \textbf{2.03} & \textbf{2.33} & \textbf{3.03} & 2.50 & 2.87 & \textbf{3.07} \\
3 & \textbf{2.00} & \textbf{2.00} & 2.90 & 2.40 & 2.43 & 2.80 & \textbf{2.00} & \textbf{2.00} & \textbf{2.03} & \textbf{2.00} & \textbf{2.00} & \textbf{2.13} \\
4 & \textbf{1.60} & \textbf{2.00} & 2.20 & 2.07 & 2.07 & 2.20 & \textbf{1.67} & \textbf{1.93} & \textbf{2.00} & 1.97 & \textbf{2.00} & \textbf{2.00} \\
\bottomrule
\end{tabular}
\caption{Maximum Queue Depth by $d$ and Priority with $\lambda=0.75$}
\label{tab:priority_max}
\end{table}

\begin{table}[h!]
\begin{tabular}{l|rrrrr}
\toprule
 $d$& Baseline & Lag=2000 & Lag=10000 & Fuzz=2 & Fuzz=10 \\
\midrule
1 & 3.02 & 2.97 & 2.99 & 3.00 & 2.98 \\
2 & 1.31 & 1.67 & 2.45 & 1.74 & 2.77 \\
3 & 1.11 & 1.60 & 2.79 & 1.59 & 2.70 \\
4 & 0.98 & 1.64 & 3.23 & 1.51 & 2.66 \\
\bottomrule
\end{tabular}
\caption{Average Queue Depth (Error Models) with $\lambda=0.75$}
\label{tab:error_avg}
\end{table}

\begin{table}

\begin{tabular}{l|rrrrr}
\toprule
$d$ & Baseline & Lag=2000 & Lag=10000 & Fuzz=2 & Fuzz=10 \\
\midrule
1 & 20.00 & 30.00 & 24.00 & 27.00 & 19.00 \\
2 & 4.00 & 8.00 & 15.00 & 7.00 & 14.00 \\
3 & 3.00 & 7.00 & 17.00 & 6.00 & 13.00 \\
4 & 3.00 & 9.00 & 17.00 & 5.00 & 12.00 \\
\bottomrule
\end{tabular}
\caption{Maximum Queue Depth (Error Models) with $\lambda=0.75$}
\label{tab:error_max}
\end{table}

\end{document}